\documentclass[preprint,11p]{elsarticle}

\usepackage{amssymb}
\usepackage{amsmath}

\usepackage{natbib}
\biboptions{numbers,sort&compress}

\usepackage{subfigure}
\usepackage{graphicx}
\usepackage{latexsym}
\usepackage{bm} 
\usepackage{amsfonts}
\usepackage{soul,color,xcolor} 
\usepackage{stfloats} 
\usepackage{booktabs} 
\usepackage{tabularx}
\usepackage{tabu}
\usepackage{multirow}
\usepackage{url}
\usepackage{geometry}
\usepackage{threeparttable} 
\usepackage[colorlinks=false, linktocpage=true]{hyperref}
\usepackage{algorithm}
\usepackage{algorithmic}

\graphicspath{{Figures/}} 

\journal{Elsevier}

\begin{document}
	
	\begin{frontmatter}
		
		
		
		\title{Optimal battery thermal management for electric vehicles with battery degradation minimization}
            
		

  		\author{Yue Wu$^{1}$, 
		Zhiwu Huang$^{1}$,
		Dongjun Li$^{2}$,
		Heng Li$^{3*}$,
		Jun Peng$^{3}$,
            Daniel Stroe$^{4}$,
		and Ziyou Song$^{2,**}$ \\
            Corresponding author: Heng Li (liheng@csu.edu.cn) and Ziyou Song (ziyou@nus.edu.sg)}
		
		\address{
			{$^{1}$ School of Automation, Central South University, Changsha 410075, China \\}
			{$^{2}$ Department of Mechanical Engineering, National University of Singapore, Singapore 117575, Singapore\\}
			{$^{3}$ School of Computer Science and Engineering, Central South University, Changsha 410075, China\\}
			{$^{4}$ Department of Energy Technology, Aalborg University, Aalborg 9220, Denmark\\}
		}
		
		\begin{abstract}
			The control of a battery thermal management system (BTMS) is essential for the thermal safety, energy efficiency, and durability of electric vehicles (EVs) in hot weather. To address the battery cooling optimization problem, this paper utilizes dynamic programming (DP) to develop an online rule-based control strategy. Firstly, an electrical-thermal-aging model of the $\rm LiFePO_4$ battery pack is established. A control-oriented onboard BTMS model is proposed and verified under different speed profiles and temperatures. Then in the DP framework, a cost function consisting of battery aging cost and cooling-induced electricity cost is minimized to obtain the optimal compressor power. By exacting three rules "fast cooling, slow cooling, and temperature-maintaining" from the DP result, a near-optimal rule-based cooling strategy, which uses as much regenerative energy as possible to cool the battery pack, is proposed for online execution. Simulation results show that the proposed online strategy can dramatically improve the driving economy and reduce battery degradation under diverse operation conditions, achieving less than a 3\% difference in battery loss compared to the offline DP. Recommendations regarding battery cooling under different real-world cases are finally provided.
		\end{abstract}
		
	\end{frontmatter}
	
 
 
	\section{Introduction}
	%
	%
	%
	%
	
	Benefiting from zero-emission and low operation cost features, electric vehicles (EVs) powered by Li-ion batteries have an increasing penetration rate in the automotive market. However, battery overheating or even thermal runaway \cite{feng2018thermal} still hinders consumer acceptance of EVs, especially in those areas with hot weather \cite{deng2018effects}. In this context, battery cooling is of vital importance to ensure thermal safety and further improve the driving economy \cite{akinlabi2020configuration}. Battery temperature is actively controlled by the battery thermal management system (BTMS) \cite{mali2021review}, which requires careful structure designs \cite{chen2017configuration,batteries9020134} to improve cooling efficiency and also, well-designed battery cooling control strategies to realize real-time, efficient, and energy-saving cooling performance \cite{dai2021advanced}. The BTMS should be able to regulate the battery temperature and maintain it within a safe range to extend the battery life and improve vehicle performance.


	Generally, BTMS can be categorized as active, passive, and hybrid systems \cite{ghaeminezhad2022review}. Active BTMSs require an additional energy source (a fan, blower, or pump) to exchange the heat between the battery pack and the thermal conductive media, e.g., air flow or liquid flow. When compared to passive and hybrid BTMSs, active BTMSs well balance the complexity of the cooling structure and cooling performance and are therefore widely applied in commercial EVs \cite{kim2019review}. Finite element analysis and computational fluid dynamics are commonly used for precise modeling \cite{wang2021cooling,lee2022investigation} and structure optimization \cite{wang2019optimization,yang2023structure} of active BTMSs. These efforts have contributed significantly to the improvement of heat exchange efficiency and the design of practical BTMSs. With a well-designed active BTMS, meticulous control is critical for BTMS operation where two main issues need to be addressed: (1) A control-oriented model with adequate accuracy and acceptable complexity is required since BTMS control involves complex refrigerant phase change and coupled electrical-thermal processes \cite{lu2020research}; and (2) Considering the coupled electrical-thermal-aging characteristics of batteries \cite{hu2020optimal}, operation optimization for battery cooling is needed to simultaneously reduce battery aging \cite{wang2011cycle} and cooling energy consumption from battery pack itself \cite{jeffers2015climate,hu2020battery}.
	
	Extensive studies about control-oriented BTMS modeling have been conducted in recent years. Amini et.al \cite{amini2020hierarchical} considered a simple constant ratio between the battery cooling rate and BTMS power, and they further developed a polynomial fitting model with acceptable accuracy, where the temperature dynamics of battery, cabin, and evaporator are involved \cite{amini2019cabin}. Similar work can also be found in \cite{zhao2021two}. To build more accurate models, Park et.al \cite{park2020computationally} further established a detailed control-oriented BTMS model considering different mass flow rates of the R134a refrigerant and pressure ratios. Kuang et.al \cite{kuang2020research} considered a high-fidelity heat transfer process and built models for the compressor, condenser, evaporator, chiller, and expansion valve.
    
	Regarding the battery cooling strategy, there have been numerous studies that can be classified into offline optimization methods \cite{bauer2014thermal, kuang2020research, zhao2021two, zhao2023two} and online optimization methods, based on, for example, look-up tables \cite{bauer2014thermal}, fuzzy PID \cite{zhao2023two}, and model predictive control (MPC) \cite{amini2020hierarchical, amini2019cabin, park2020computationally, park2021model, broatch2022leveraging, zhao2021two, xie2020mpc}.
 
	In terms of offline cooling strategies, a hierarchical and iterative dynamic programming (HIDP) scheme \cite{zhao2021two} was proposed to provide the optimal battery temperature trajectory so that an MPC can track and minimize cooling power. Similarly, Zhao et.al \cite{zhao2023two} proposed a fuzzy PID to track the optimal temperature trajectories from dynamic programming (DP), where the optimal battery target temperature is calculated considering weather conditions, passenger characteristics, and battery conditions. Kuang et. al \cite{kuang2020research} proposed a genetic algorithm method to explicitly minimize the battery capacity loss (i.e., battery degradation) and cooling power, where different weight coefficients were discussed carefully. Bauer et.al \cite{bauer2014thermal} developed an offline Pontryagin’s maximum principle (PMP)-based battery cooling strategy to minimize the consumed battery energy and battery temperature tracking error (then a look-up table based on the PMP is designed for online execution). The above offline battery cooling optimization methods all require EVs' entire traction power/velocity profile as a priori and suffer from heavy computation burdens, which are difficult to be used in practice.
 
	To realize online cooling optimization, MPC can be a promising online optimization method leveraging future information \cite{wu2023spatial} and has been widely investigated. A hierarchical MPC framework is proposed for cabin/battery thermal management optimization in EV \cite{amini2020hierarchical} and hybrid electric vehicles (HEVs) \cite{amini2019cabin}. Park et.al \cite{park2020computationally, park2021model} developed a stochastic MPC to handle uncertainties of future information by designing a stochastic model of future heat generation. In these studies, the cooling power and battery temperature tracking error are optimized. Similarly, a probability matrix based on the Markov chain is proposed on a commuting road \cite{broatch2022leveraging} to estimate the future battery temperature for the MPC execution, to minimize the cooling cost and battery temperature tracking error. Xie et.al \cite{xie2020mpc} proposed a battery cooling MPC method to optimize the battery temperature tracking error, coolant flow rate, and the change rate of coolant flow.
	
	Despite the above-related studies, there are still several issues to be resolved. To the best of our knowledge, (1) A control-oriented BTMS model that is effective under different working conditions, e.g., EV velocity, environment temperature, and BTMS parameters, is still missing in the available literature; (2) An online battery cooling optimization method that minimizes the battery capacity degradation has not been reported yet; (3) From the perspective of battery degradation and driving economy, how to cool the battery under different conditions is still an open problem.

	To address the challenges, this paper presents a comprehensive investigation of EV battery cooling optimization based on dynamic programming. Firstly, a control-oriented BTMS model is established and verified under different working conditions. Then an offline DP-based battery cooling strategy is proposed to minimize the costs of battery capacity degradation and battery cooling. By analyzing the DP results under various driving conditions (urban, suburban, and highway), three rules are extracted to construct a near-optimal rule-based cooling strategy for online implementation. Finally, a quantitative analysis and comparison with existing studies are conducted to indicate the superiority and optimality of the proposed online strategy, together with a comprehensive analysis of the driving economy under versatile conditions. The contributions of this paper can be summarized below.
	
	1. A control-oriented BTMS model, which has high accuracy under wide ranges of EV velocity (0-120km/h), environmental temperature (26-40 $^{\circ}$C), and mass ﬂow rate of the coolant (0.14-0.22kg/s), is established through orthogonal experiments and data fitting.
	
	2. The proposed online cooling strategy, which can minimize battery capacity loss and cooling energy by leveraging regenerative power, is verified to achieve near-optimal performance with ease of implementation in EV applications when compared to DP results.
	
	3. A comprehensive and quantitative driving economy analysis is presented to explore the performance of the proposed cooling strategy under versatile operating conditions, i.e., driving conditions, environmental temperatures, and BTMS parameters. Recommendations for battery cooling are then provided.
	
	The remainder of this paper is organized as follows. In Section \ref{model}, detailed modeling of the battery pack and BTMS is introduced. Offline DP optimization for battery cooling together with the results are illustrated in Section \ref{DP}. Analysis of the DP results and the rule-based cooling strategy are proposed in Section \ref{rule}. Section \ref{result} presents the detailed results and analysis. Conclusions are drawn in Section \ref{conclusion}.
	
	\section{System Modeling \label{model}}
	
\begin{figure}[!h]
		\centering
		\includegraphics[width=8cm]{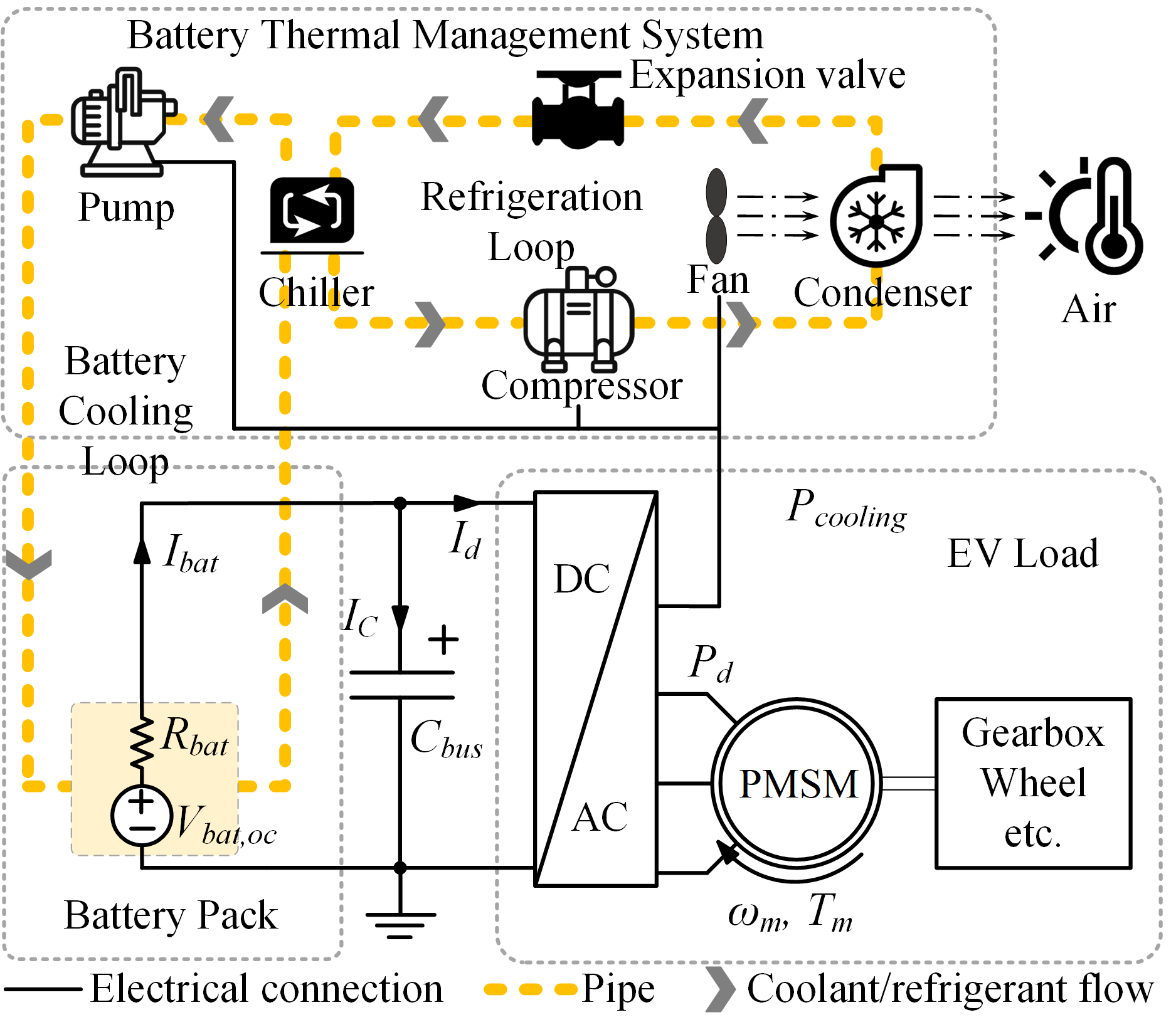} 
		\caption{Battery energy storage system, battery thermal management system, and traction load of the electric vehicle.}
		\label{fig1_config}
	\end{figure}

	As shown in Fig. \ref{fig1_config}, the studied system includes a battery pack as the energy storage system, and the auxiliary load corresponding to the BTMS, which consists of two thermal loops, i.e., the battery cooling loop and the refrigeration loop. The battery cooling power $P_{cooling}$ comes from the electricity consumption of the compressor, the pump, and the fan.
	
	\subsection{EV model}
	
	The power balance between the battery pack, the BTMS, and the traction system is given as
	
	\begin{equation}
		\begin{aligned}
			\label{eq:powerbalance}
			(P_{d} + P_{cooling})/\eta_{dcac} = P_{bat}
		\end{aligned}
	\end{equation}
	where $P_{d}$ is the traction/braking power, $P_{cooling}$ is the total power of the compressor, the pump, and the fan, and $P_{bat}$ is the power of the battery pack. Detailed modeling and parameters of the DC/AC inverter, motor, powertrain, and vehicle dynamics can be found in our previous work \cite{wu2022hierarchical}.
	
	\subsection{Battery electrical model}
	
	The $Rint$ model is considered for modeling the dynamics of the battery pack. The voltage, power, and state-of-charge (SoC) dynamics are given below:
	
	\begin{equation}
		\label{eq:Vbat}
		V_{bat} = V_{bat,oc} - I_{bat}R_{bat}
	\end{equation}
	
	\begin{equation}
		\label{eq:Pbat}
		P_{bat} = V_{bat,oc}I_{bat} - I^2_{bat}R_{bat}
	\end{equation}	
	
	\begin{equation}
		\label{eq:SoCbat}
		{SoC}_{bat,k+1} = {SoC}_{bat,k} - \frac{I_{bat,k}T_{s}}{3600Q_{bat}},
	\end{equation}
	where $V_{bat,oc}$ and $V_{bat}$ represent the open-circuit voltage and terminal voltage, $R_{bat}$ is the internal resistance, $V_{bat,oc}$ is a function of battery SoC, given in \cite{song2018component}, $R_{bat}$ is a function of charging/discharging, SoC, and temperature, provided in \cite{song2014multi}, and $Q_{bat}$ is the capacity of the battery pack.
	Other basic parameters of the $\rm LiFePO_4$ battery cell considered in this work are given in Table \ref{tab1}. In the studied EV, 250 60Ah cells form a 120Ah - 412V battery pack \cite{huang2022sizing} (125 in series, 2 in parallel).
	
	\begin{table}[!h] 
		\caption{Parameters of the Li-ion battery}
		\label{tab1}
		\centering
		\begin{tabular}{lc}
			\toprule
			Parameter							&Battery (cell) 			\\
			\midrule
			Capacity							&60Ah						\\					
			Nominal voltage 					&3.3V						\\
			Stored energy						&192Wh						\\
			Mass								&2.5kg						\\
			SoC range							&[5\%,100\%]				\\
			\bottomrule
		\end{tabular}
	\end{table}
	
	\subsection{Battery aging model}
	
	A dynamic capacity degradation model for $\rm LiFePO_4$ batteries \cite{wang2011cycle} is utilized to quantify the battery degradation. Through experimental testing and parameter calibration from 5$^{\circ}$C to 45$^{\circ}$C \cite{song2015optimization}, the discrete form of this model is provided as 
	
	\begin{equation}	
		\label{eq:Qlossk}
		\Delta Q_{loss,k} = 9.78 \times 10^{-4} \frac{|I_{bat,k}|T_{s}}{3600}  e^{(\frac{-15162+1516C_{rate,k}}{0.849RT_{bat,k}})} Q_{loss,k-1}^{-0.1779}
	\end{equation}
    where $\Delta Q_{loss,k}$ is the instantaneous capacity loss (\%), $R$ is the gas constant 8.314J/(mol$\cdot$K), $T_{bat,k}$ is the battery absolute temperature (K), $C_{rate,k}$ is the current rate (i.e., $|I_{bat}|/Q_{bat}$), and $T_s$ denotes the sampling time. Therefore, the incremental capacity reduction $\Delta Q_{loss,k}$ is a non-linear function of $I_{bat,k}$ and $T_{bat,k}$, and it will be directly incorporated in the cost function in this paper.
	
	\subsection{Battery thermal model}
	
	Heat generation of the battery pack mainly consists of resistance joule heat and reversible entropy heat \cite{song2015heating}, which are the first term and the second term in Eq. (\ref{eq:qbat}).
	
	\begin{equation}
		\label{eq:qbat}
		\dot Q_{gen} = I^2_{bat}R_{bat} + I_{bat}T_{bat} \frac{dV_{bat}}{dT_{bat}}
	\end{equation}
	where $\frac{dV_{bat}}{dT_{bat}}$ is derived through the entropy potential calibration of the $\rm LiFePO_4$ cell, the detailed results can be found in \cite{song2014multi} (positive current denotes discharging). For simplicity, all battery cells are assumed to have the same electrical and thermal parameters, i.e., no temperature variation is considered among cells. Moreover, this study does not distinguish the core temperature and surface temperature of the battery cell, the core temperature can be estimated using the surface temperature as described in \cite{liu2022online}. Hence, the battery temperature dynamics can be derived, as shown below \cite{song2015heating}.
	
	\begin{equation}
		\label{eq:Tbat}
		\frac{dT_{bat} }{dt} = \frac{\dot  Q_{gen} - \dot Q_{cool}}{N_{cell}C_{cell}}
	\end{equation}
	where $N_{cell} = 250$ is the total number of battery cells in the pack, $C_{cell}$ is the thermal capacity of the battery cell, i.e., 2299J$/^{\circ}$C, and $\dot Q_{cool}$ is the battery cooling rate, determined by the BTMS.
	
\begin{figure}[!t]
		\centering
		\includegraphics[width=8cm]{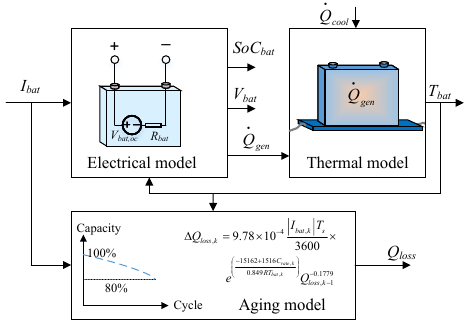} 
		\caption{Coupling relationship between battery electrical, thermal, and aging models.}
		\label{fig2_coupling}
	\end{figure}

	The electrical-thermal-aging coupling relationship of the Li-ion battery is shown in Fig. \ref{fig2_coupling}. The battery current $I_{bat}$ is the input of these coupled models, then battery states such as SoC, terminal voltage, heat generation, temperature, and capacity loss are calculated. Battery temperature $T_{bat}$ will be fed back to update the electrical and aging models iteratively and is determined by the environment temperature, heat generation, and cooling rate.
	
	\subsection{Battery thermal management system model}
	
\begin{figure}[!h]
		\centering
		\includegraphics[width=8cm]{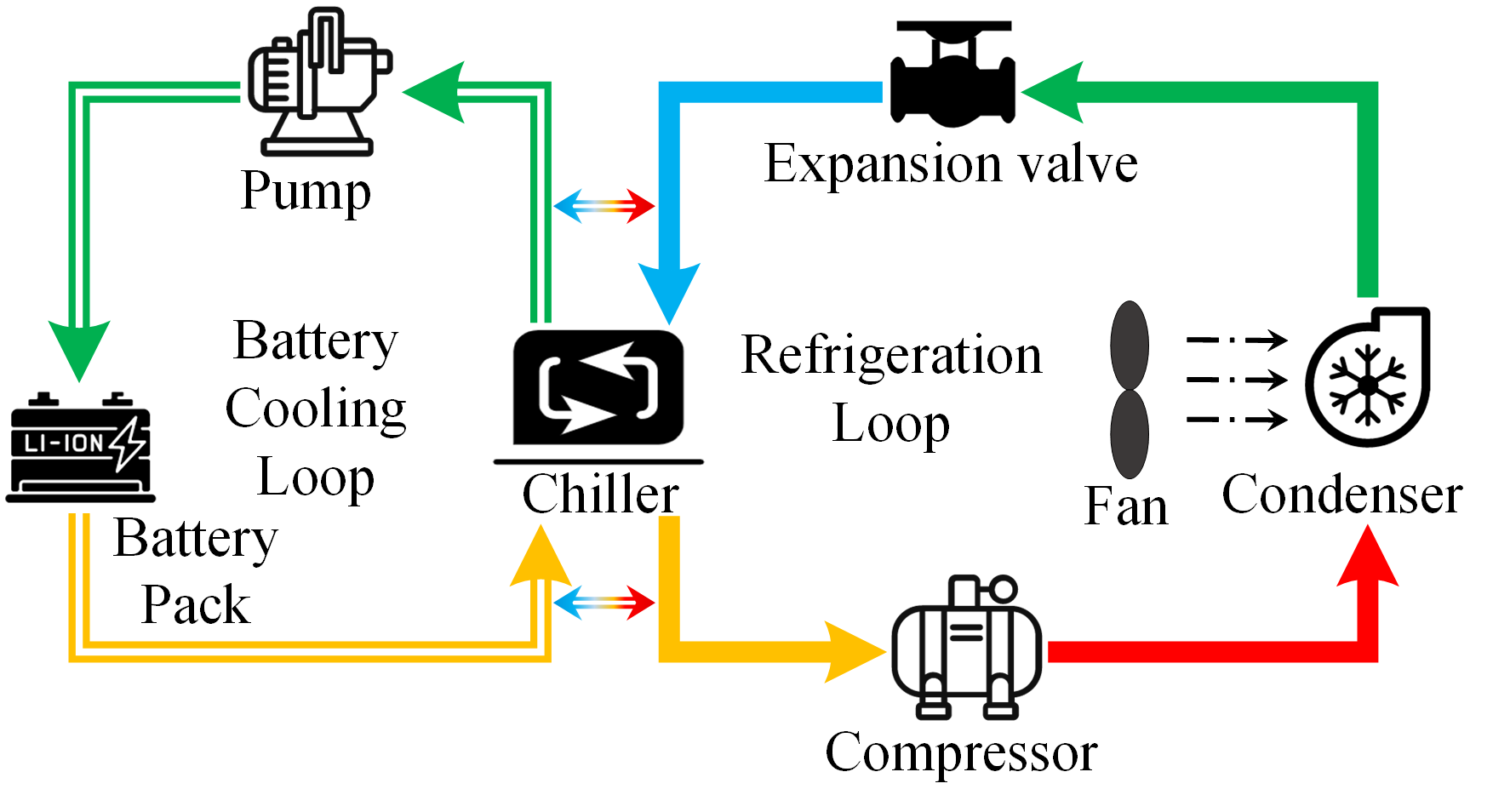}
		\caption{Battery thermal management system. Red, orange, green, and blue represent the temperatures from high to low.}
		\label{fig3_ACframework}
	\end{figure}

	Fig. \ref{fig3_ACframework} shows the framework and schematic of the studied BTMS, where the colored lines highlight the two thermal loops. The solid line represents the refrigeration loop, and the double solid line represents the battery cooling loop. For the refrigeration loop, driven by the compressor, the refrigerant undergoes several procedures through the compressor, condenser, expansion valve, and Chiller. It turns into high-pressure superheated vapor from low-pressure superheated vapor through compression, then dissipates heat to ambient air through the condenser with a new state of high-pressure cooled liquid. After the expansion valve, the refrigerant turns into low-pressure mixed liquid and vapor, while the temperature drops due to liquid vaporization. Finally, the mixed refrigerant liquid and vapor absorb heat through the chiller.

    The above processes form a compressed refrigeration cycle of the air-conditioning system \cite{wang2022cooling, wang2020eco}. For the battery cooling loop, driven by the pump, the coolant flows into the battery pack to absorb heat generated by the battery pack. Then the heat exchange process between the coolant and refrigerant will occur while the coolant flows into the Chiller. To sum up, the studied BTMS model contains three sub-models: a sub-model describes the relationship between cooling power and consumed energy of BTMS (sub-model I), a sub-model describing the heat exchange between the battery pack and coolant (sub-model II), and a sub-model describing the heat exchange between the coolant and refrigerant in Chiller (sub-model III).	

     For sub-model I, the total electricity consumption comes from three parts, i.e., pump, fan, and compressor. Considering that the power levels of the pump ($<$ 50W) and the fan ($\sim$100-200W) are much lower than the compressor (500-4500W), and the fan is also operated for the cooling loop of the electric drive (detailed vehicle thermal management system framework can be found in \cite{li2021innovative}), thus only $P_{comp}$ is considered as the control variable in the studied BTMS. The fan and the pump are set to be always on. Assuming a constant power of 200W is considered for the pump and the fan, then we have 
	
	\begin{equation}
		\label{eq:Pcooling}
		P_{cooling} = P_{comp} + 200.
	\end{equation}
	
\begin{figure}[!h]
		\centering
		\includegraphics[width=8cm]{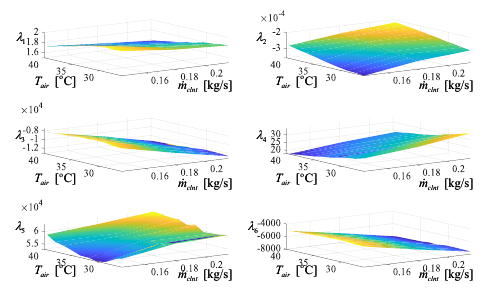}
		\caption{Model fitted coefficients of the BTMS model $\lambda_i$.}
		\label{fig4_lamda}
	\end{figure}

	As mentioned above, the compression refrigeration process is a highly nonlinear process, which requires quantities of complicated equations to describe. To reduce calculation burden and model complexity, a model describing the relationship between cooling rate and power consumed by the AC system is fitted by orthogonal simulation experiment data. Data was generated by a vehicle AC system model built into KULI software \cite{li2021innovative, qi2007analysis,wang2022cooling, wang2020eco}, this model was well validated by experimental vehicle data. The model is given as
	
	\begin{equation}
		\begin{aligned}
			\label{eq:Q_cool}
			\dot{Q}_{cool} = \lambda_1 P_{comp} + \lambda_2 P^2_{comp} + \lambda_3 T_{clnt,out} + \lambda_4 T_{air} \dot{m}_{air} \\+  \lambda_5 T_{clnt,out} \dot{m}_{clnt} + \lambda_6
		\end{aligned}
	\end{equation}
 
 	\begin{equation}
		\label{eq:m_air}
		\dot{m}_{air} = 0.07065 + 0.00606v
	\end{equation}
	where $\dot{Q}_{cool}$ represents the supplied cooling rate (power) to the battery pack from the BTMS, $P_{comp}$ is the compressor electrical power, $T_{clnt,out}$ is the outlet temperature of the coolant, $T_{air}$ is the environment temperature, $\dot{m}_{air}$ the total air mass flow through the condenser, $\dot{m}_{clnt}$ is the mass flow rate of coolant, $v$ is the EV velocity (km/h), and $\lambda_i$ are different coefficients determined by $T_{air}$ and $\dot{m}_{clnt}$. Eq.(\ref{eq:m_air}) describes the relationship between the air mass flow rate and the EV velocity. In addition, all coefficients $\lambda_i$ are functions of $T_{air}$ and $\dot{m}_{clnt}$, as shown in Fig. \ref{fig4_lamda}. With these fitted coefficients, more than 80\% and 97\% of the model/experiment data have an error of less than 5\% and 10\%, respectively, see Fig. \ref{fig5_error}. More details of the BTMS orthogonal experiment and $\lambda_i$ values are given in Appendix.
	
	\begin{figure}[!h]
		\centering
		\includegraphics[width=8cm]{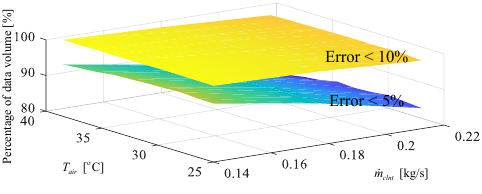}
		\caption{Percentage of data with model error less than 5\% and 10\%.}
		\label{fig5_error}
	\end{figure}
	
	The working ranges of the key variables are determined by application scenarios and operation range of each component \cite{wang2022cooling,li2021innovative,wang2019performance}, see Table \ref{tab_range}. $P_{comp}$ is the control variable with operation range [500, 4500W]. When $P_{comp}$ is too low, i.e., less than 500W, the compressor has a low rotation speed (less than 500rpm), thereby failing to repress refrigerant and negatively impacting the compressor. As a result, we assume $\dot{Q}_{cool} = 0$ when $P_{comp} <500$W.

    \begin{table}[!h] 
		\caption{Application scope of BTMS orthogonal experiment}
		\label{tab_range}
		\centering
		\resizebox{8cm}{!}{
			\begin{tabular}{lll}
				\toprule
				Input variables							                &Value 			          &Unit   \\
				\midrule
				Outlet temperature of coolant	$T_{clnt,out}$			&25-39	              &$^{\circ}$C   \\	
				Volumetric flow rate of coolant	$\dot{q}_{clnt}$		&8-12	                  &L/min   \\				
				Ambient air temperature $T_{air}$	                    &26-39.5		          &$^{\circ}$C   \\
				EV velocity $v$	                                    &0-120		          &km/h   \\
				Compressor power $P_{comp}$                 	&500-4500	          &W   \\
				\bottomrule
		\end{tabular}}
	\end{table}
	
	The following Eqs. (\ref{eq:T_clnt_out}) and (\ref{eq:T_clnt_in}) describe the heat exchange process of sub-models II and III of BTMS.
	
	\begin{equation}
		\label{eq:T_clnt_out}
		T_{clnt,out} = (T_{clnt,in} - T_{bat})e^{\frac{-h_{bat}A_{bat}}{\dot{m}_{clnt}{c}_{clnt}}} + T_{bat}
	\end{equation}
	
	\begin{equation}
		\label{eq:T_clnt_in}
		T_{clnt,in} = T_{clnt,out} - \frac{\dot{Q}_{cool}}{\dot{m}_{clnt}{c}_{clnt}} 
	\end{equation}
	where $T_{clnt,in}$ is the temperature of the coolant inlet, $h_{bat}$ represents the heat transfer coefficient between the battery and the coolant, $A_{bat}$ is the heat transfer area between the battery and the coolant, $c_{clnt}$ denotes the specific heat capacity of the coolant. All the constant parameters of sub-model II and III are listed in Table \ref{tab2}. 
	
	\begin{table}[!h] 
		\caption{Parameter settings of the BTMS}
		\label{tab2}
		\centering
		\begin{threeparttable}
			\resizebox{8cm}{!}{	
				\begin{tabular}{lcc}
					\toprule
					Description and symbol							&Value 	                                    &Unit		\\
					\midrule
					Specific heat capacity of coolant $c_{clnt}$	&3330  \cite{park2020computationally}		&J/kg/$^{\circ}$C \\
					Heat transfer coefficient between battery/coolant &300 \cite{park2020computationally}       &W/m$^{2}/^{\circ}$C \\ 
					Mass flow rate of the coolant $\dot{m}_{clnt}$	&0.18                                       & kg/s	\\
					Heat transfer area between battery/coolant $A_{bat}$ &3.1                                   &m$^{2}$	\\
					Environment air temperature $T_{air}$	&33                                                 &$^{\circ}$C 	\\
					Initial temperatures of the coolant inlet $T_{clnt,in}$	&33                                 &$^{\circ}$C 	\\
					Initial temperatures of the coolant outlet $T_{clnt,out}$	&33                             &$^{\circ}$C		\\
					\bottomrule
			\end{tabular}}
		\end{threeparttable}
	\end{table}
	
	\section{Offline optimization of battery thermal management \label{DP}}
	
	This section illustrates the offline optimization of battery cooling using DP under typical driving conditions, and the DP results may provide some insightful guidance to design online battery cooling strategies.
	
	\subsection{Dynamic programming development}
	
\begin{figure}[!h]
		\centering
		\includegraphics[width=8cm]{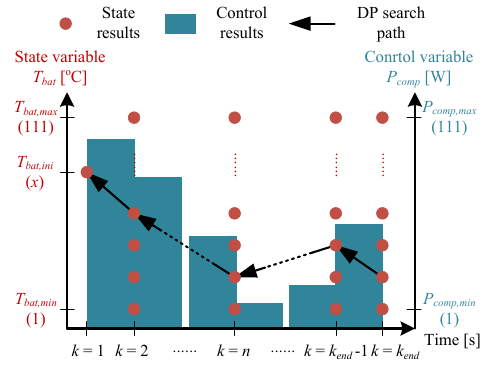}
		\caption{Principle of DP search procedure.}
		\label{fig6_DP}
	\end{figure}

	First, we set a target battery temperature of 25$^{\circ}$C in the cooling process under hot weather conditions, as generally, there is no need to cool the battery down to a lower temperature for most battery chemistries \cite{karimi2013thermal,ghaeminezhad2022review}, considering the tradeoff between energy consumption and battery degradation rate \cite{zhao2023two}. In the DP process, the primary state is the battery temperature $T_{bat}$, which is discretized into 111 states (with the temperature resolution of $\sim$0.1$^{\circ}$C) in the range of [$T_{bat,min}$, $T_{bat,max}$], where $T_{bat,min} = 24 ^{\circ}$C, $T_{bat,max} = T_{air}+2^{\circ}$C. Temperature margins of 1-2 degrees are set to ensure all temperatures around 25$^{\circ}$C and $T_{air}$ can be considered, i.e., the optimal control variable is available for the query if the battery temperature is slightly below 25$^{\circ}$C or above $T_{air}$ in the forward execution. The control variable is the compressor power $P_{comp}$, which is also divided into 111 discrete states. The backward search procedure of DP on the time domain is shown in Fig. \ref{fig6_DP}, when all discrete states and actions are traversed, an offline table of the Value function is established for the forward execution \cite{bellman1966dynamic}.

	Battery cooling may result in reduced battery degradation but bring an increasing energy consumption. Thus the total cost including the battery capacity loss (aging) and the accumulated energy consumption for cooling under the entire driving cycle needs to be minimized by the DP, as shown below. 
	
	\begin{equation}
		\begin{aligned}
			\label{eq:DP}
			\text{Minimize}\big[\sum_{k=1}^{k_{end}}(\frac{Q_{bat}V_{bat}price_{bat}}{1000} \times \frac{\Delta Q_{loss,k}T_{s}}{0.2} + price_{ele}P_{cooling,k}T_{s})\big],	
		\end{aligned}
	\end{equation}
	where the first item is the battery capacity degradation cost, and the second item is the electricity cost associated with cooling the battery. Furthermore, $price_{bat}$ is the current battery price, i.e., 150USD/kWh \cite{BatPrice}, and the constant 0.2 denotes the maximum allowable battery capacity loss of 20\% in EVs (the initial capacity of the battery pack is normalized to 1 \cite{song2014multi}), $\Delta Q_{loss,k}$ is given in Eq. (\ref{eq:Qlossk}), $price_{ele}$ is the electricity price, i.e., 0.1USD/kWh, $P_{cooling,k}$ is mainly determined by $P_{comp,k}$, and $T_{s}$ is set to 1s. This cost function is calculated considering 5 initial capacity losses 0.01\% (a very small number that is close to 0 to avoid numerical issues and utilize the model successfully), 5\%, 10\%, 15\%, and 20\% to assess the average cost over the entire battery life \cite{wu2022hierarchical}. The constraints involved in the DP optimization are given below:
	
	\begin{equation}
		\label{eq:cons}
		\begin{cases}
			0 \leq P_{comp} \leq 4500\text{W}\\
			T_{bat,target} = 25^{\circ}\text{C}\\
			I_{bat,min} \leq I_{bat} \leq I_{bat,max}\\
			SoC_{bat,min} \leq SoC_{bat} \leq SoC_{bat,max}\\
		\end{cases}
	\end{equation}
	for the second constraint, when the battery temperature is lower than or equal to 25$^{\circ}$C, the searching range of $P_{comp}$ will be modified to [0, 0].
	
	\subsection{Analysis of dynamic programming results \label{DP results}}
	
\begin{figure}[!h]
		\centering
		\includegraphics[width=8cm]{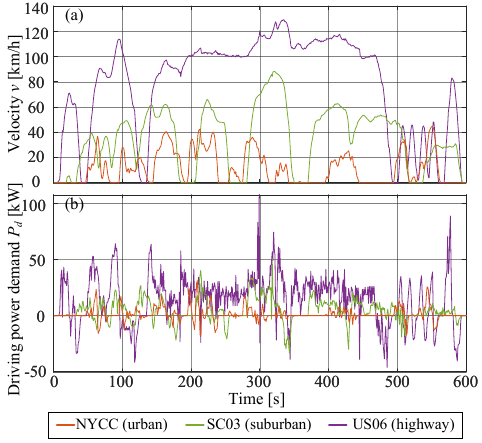}
		\caption{Three representative driving cycles: NYCC for urban, SC03 for suburban, and US06 for highway. (a) velocity, and (b) power demand.}
		\label{fig7_AllSpeed}
	\end{figure}

	This study investigates three representative driving cycles, i.e., NYCC, SC03, and US06, representing typical urban, suburban, and highway conditions, respectively. As shown in Fig. \ref{fig7_AllSpeed} (a), the three driving cycles have the same time duration of 600s, with average velocities of 11.40km/h, 34.58km/h, and 77.36km/h and average distances of 1.90km, 5.76km, and 12.89km, respectively. Fig. \ref{fig7_AllSpeed} (b) shows the driving power demand $P_d$ calculated based on the EV dynamic model.
	
	The initial battery SoC is 95\%, thus the DP is performed on the repeated driving cycles to consume the battery until its SoC is below 10\%. Due to the different distances, NYCC is repeated 165 times, while SC03 and US06 are repeated 55 and 18 times, respectively, leading to the total distances of 314km, 317km, and 232km for those three driving cycles.
	
\begin{figure*}[!h]
		\centering
		\includegraphics[width=16cm]{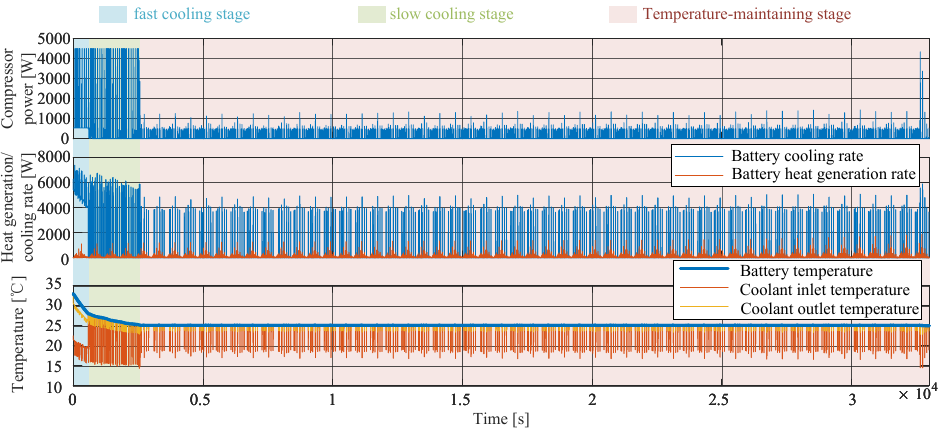}
		\caption{Offline DP results of compressor power, battery heating/cooling rate, and temperature under 55 SC03 driving cycles (33000s).}
		\label{fig8_DPSC03}
	\end{figure*}

	Taking SC03 as an example (the detailed DP results of NYCC and US06 are not given in this section due to the limitation of the article length), the results of compressor power, battery heat generation/cooling rate, and battery/coolant temperatures are illustrated in Fig. \ref{fig8_DPSC03}. The whole battery cooling process can be divided into three stages: the fast cooling stage from 0s to 580s, the slow cooling stage from 581s to 2600s, and the temperature-maintaining stage from 2601s to the end. 

	Specifically, in the fast cooling stage, the compressor is always working with $P_{comp}$ between 500W and 4500W. At the end of fast cooling, the battery temperature is about 28$^{\circ}$C. In this stage, the average coolant inlet temperature is 10$^{\circ}$C lower than the battery temperature, and the average coolant inlet temperature is 2.25$^{\circ}$C lower than the battery temperature. In the slow cooling stage, the compressor will not continuously work, while the compressor power can still reach the maximum value of 4500W. At the 2600s, the battery temperature drops to about 25$^{\circ}$C. In the slow cooling and temperature-maintaining stages, the temperatures of coolant at the inlet and outlet can be equal to battery temperature sometimes, since the compressor only works intermittently. When the compressor doesn't work, the temperatures of coolant at the inlet and outlet will gradually increase through heat exchange with the battery. For the fast cooling stage, the temperatures of coolant at the inlet and outlet are always lower than the battery with continuous compressor operation. When it goes into the temperature-maintaining stage, the compressor power is between 0 and a low value to maintain the battery temperature around 25$^{\circ}$C.

	\section{Online battery cooling strategy \label{rule}}
	
\begin{figure}[!h]
		\centering
		\includegraphics[width=8cm]{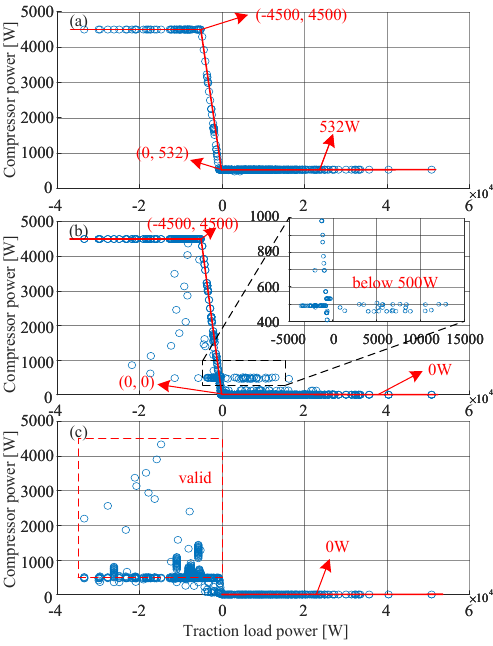}
		\caption{Optimal rules from the relationship between the compressor power $P_{comp}$ and traction power $P_d$ in  DP results under SC03. (a) is the fast cooling stage, (b) is the slow cooling stage, and (c) is the temperature-maintaining stage.}
		\label{fig11_rule}
	\end{figure}

	In this section, an online rule-based cooling strategy is proposed according to the DP results to achieve near-optimal performance with a low computational cost for practical implementations. To explore how DP balances the energy consumption and battery degradation in the cooling process, we carefully analyze the DP results, especially the underlying insights, i.e., the relationship between the traction power $P_d$ and the compressor power $P_{comp}$ under different stages. Specifically, as shown in Fig. \ref{fig11_rule} (a), in the fast cooling stage, the compressor is always cooling effectively ($>$ 500W). When the EV is in traction mode ($P_d>0$), the compressor power is mainly constant at $\sim$532W. When the EV is in regenerative braking mode ($P_d<0$), the compressor power equals the regenerative power, and the part that exceeds 4500W is limited and used for charging the battery. As a result, the battery degradation can be remarkably reduced, as charging, especially at a high rate, will degrade the battery fast.

\begin{figure}[!h]
		\centering
		\includegraphics[width=8cm]{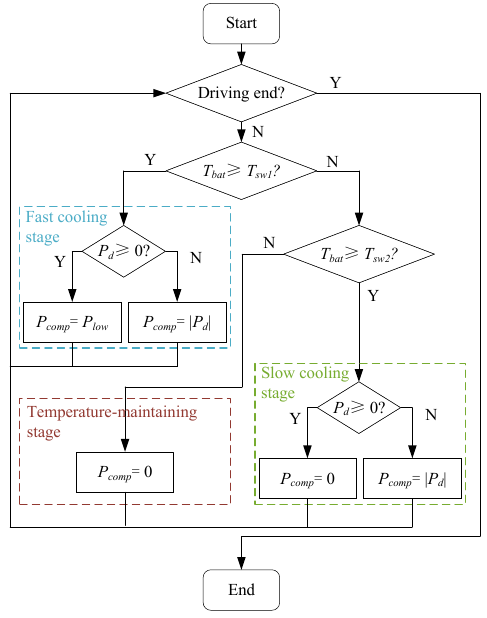}
		\caption{Flow chart of the online near-optimal battery cooling strategy.}
		\label{fig12_strategy}
	\end{figure}

	As shown in Fig. \ref{fig11_rule} (b), in the slow cooling stage, when the EV is in regenerative braking mode ($P_d<0$), the cooling rule remains the same as in the fast cooling stage. However, when the EV is in traction mode ($P_d>0$), the compressor does not function with $P_{comp}$ concentrating around 0 W, meaning that all compressor power comes from the regenerative braking power. Note that even though there are a few points around 500W due to DP grid interpolation in the forward execution, most of them are below 500W and therefore negligible.
	As shown in Fig. \ref{fig11_rule} (c), in the temperature-maintaining stage, the compressor does not function ($P_{comp} = 0$) when the EV is in traction mode ($P_d>0$). In regenerative braking mode, the compressor can hardly work at its maximum power of 4500W, and the points above 500W do not show a clear pattern. Despite this, the battery temperature will be maintained at around 25$^{\circ}$C, fluctuating in a small range of $\sim$0.1$^{\circ}$C. Note that it is hard to execute the possible rule between $P_{comp}$ and $T_{bat}$ in practice. Therefore, all these valid points will be ignored for easy implementation, i.e., $P_{comp} = 0$. When the battery temperature rises, the BTMS can switch back to the slow cooling stage and cool the battery.
    Finally, three cooling rules can be extracted for the three cooling stages, and switching conditions based on the battery temperature are added according to the DP results, as illustrated in Fig. \ref{fig12_strategy}. Note that DP also behaves similarly under NYCC and US06, all key parameters of the rule-based battery cooling strategy under NYCC, SC03, and US06 are listed in Table \ref{tab5}.
 
	\begin{table}[!h] 
		\caption{Parameters of the proposed battery cooling strategy.}
		\label{tab5}
		\centering
		\begin{threeparttable}
			\resizebox{8cm}{!}{	
				\begin{tabular}{lccc}
					\toprule
					Parameter	& $T_{sw1}$ ($^{\circ}$C)  & $T_{sw2}$ ($^{\circ}$C)	& $P_{low}$ (W)		\\
					\midrule
					NYCC		& 31	& 25  &  532		\\
					SC03		& 28 	& 25  &  532		\\
					US06		& 26	& 25  &  532		\\
					\bottomrule
			\end{tabular}}
		\end{threeparttable}
	\end{table}
	
	For the three different driving cycles, only $T_{sw1}$ is different, as high velocity will induce more heat generation in batteries, thereby requiring a longer fast cooling stage and a lower $T_{sw1}$. All $T_{sw1}$ and $T_{sw2}$ values are integers for ease of implementation, i.e., fractional digits are ignored. Here we can investigate why the battery cooling process is separated into fast and slow cooling stages by DP. The reason is that $T_{sw1}$ is a temperature threshold related to the driving conditions, i.e., when $T_{bat}>T_{sw1}$, the low cooling power $P_{low}$ can bring about a battery degradation cost reduction that is higher than electricity cost incurred by cooling, while when $T_{bat}<T_{sw1}$, the battery degradation cost reduction is less than the cooling electricity cost, as indicated in the optimal $P_{comp}$ lookup table of DP. In Table \ref{tab5}, NYCC necessitates the highest $T_{sw1}$, while US06 requires the lowest, for the reason that higher vehicle speeds result in higher average traction power $P_d$. As a result, the proportion of additional cooling power $P_{low}$ to traction power $P_d$ decreases, and the proportion of additional battery degradation from $P_{low}$ to the total power decreases as well. Therefore, a lower temperature threshold $T_{sw1}$ is desired. In Fig. \ref{fig7_AllSpeed} (b), the average traction power $P_d$ of NYCC, SC03, and US06 is 1.30, 4.43, and 14.56kW, respectively.
	
	\section{Results and Discussions \label{result}}
	
	In this section, the proposed online cooling strategy is first compared with the offline DP, existing MPC, and BTMS off (no-cooling) cases. Then, a study of how driving distance impacts the driving economy and battery degradation is conducted to investigate the superiority of the proposed cooling strategy under different driving scenarios. Finally, a sensitivity analysis of three parameters (environment temperature $T_{air}$, mass flow rate of the coolant $\dot{m}_{clnt}$, and heat transfer area between the battery and coolant $A_{bat}$) is conducted to verify the robustness of the cooling strategy, and recommendations of battery cooling under different conditions are provided.
	
	\subsection{Comparative results of battery cooling process}
	
	The proposed rule-based cooling strategy is compared with the offline DP and online MPC to validate its performance. For existing online optimization MPC methods \cite{amini2020hierarchical, amini2019cabin, park2020computationally, park2021model, broatch2022leveraging, zhao2021two, xie2020mpc}, the battery degradation is not explicitly minimized, thus we design the following cost function for the MPC, which is consistent with the above literature, as given by

        \begin{equation}
		\begin{aligned}
			\label{eq:JMPC}
			min J = \sum_{t=k}^{k+N_{p}-1}[\alpha(T_{bat}-T_{bat,target})^2 + price_{ele}P_{cooling,k}T_{s}], \\ k=0, 1, 2, ... 
		\end{aligned}
	\end{equation}
 where $N_{p}$ is the prediction horizon of the MPC, $\alpha$ is the weighting factor of the temperature tracking term, and the second term of the cost function is the battery cooling cost which is the same as DP. The prediction horizon is 10, and a perfect prediction of EV velocity/traction load power (no prediction error) is considered. Note that such prediction is almost impossible in real-world implementations; thus, the MPC performance will deteriorate \cite{wu2023spatial}.

    \begin{figure}[!h]
		\centering
		\includegraphics[width=8cm]{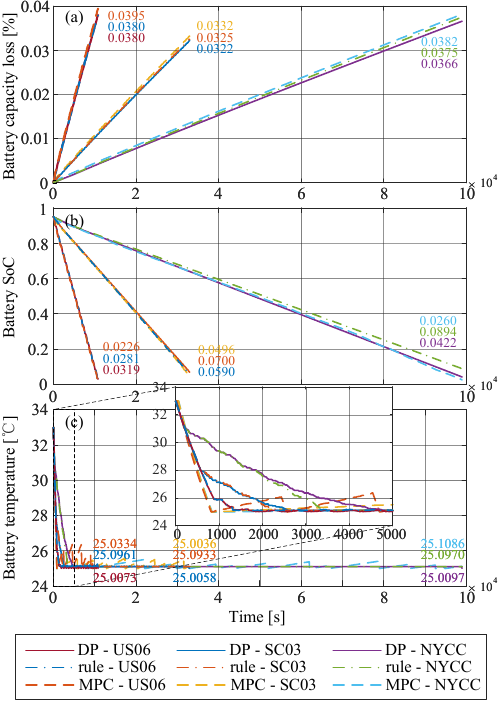}
		\caption{Comparison results of DP, MPC, and the proposed strategy in terms of (a) battery capacity loss, (b) battery SoC, and (c) battery temperature under different driving conditions.}
		\label{fig13_comparison}
	\end{figure}

    \begin{table*}[!b] 
		\caption{Numerical comparison results of the proposed online strategy, DP method, MPC, and no-cooling case.}
		\label{tab6}
		\centering
		\begin{threeparttable}
			\resizebox{17.5cm}{!}{	
				\begin{tabular}{lcccccccccccc}
					\toprule
					Index & NYCC & & & & SC03 & & & & US06 & & &	\\
					& DP & rule-based strategy & MPC & BTMS off & DP & rule-based strategy & MPC & BTMS off & DP & rule-based strategy & MPC & BTMS off	\\
					\midrule
					Battery capacity loss (\%)	& 0.0366 & 0.0375 & 0.0382 & 0.0462 & 0.0322 & 0.0325 & 0.0332 & 0.0400 & 0.0380 & 0.0380 & 0.0395 & 0.0474 \\
					Final battery SoC  			& 0.0422 & 0.0894 & 0.0260 & 0.1092 & 0.0590 & 0.0700 & 0.0496 & 0.0884 & 0.0319 & 0.0281 & 0.0226 & 0.0536 \\
					Final battery temperature ($^{\circ}$C)	& 25.0097 & 25.0970 & 25.1086 & 35.7643 & 25.0058 & 25.0933 & 25.0036 & 36.1589 & 25.0073 & 25.0961 & 25.0334 & 38.3696 \\
					Fast cooling period (s)		& [0, 250] & [0, 221] & [0, 840] & - & [0, 580] & [0, 590] & [0, 832] & - & [0, 887] & [0, 899] & [0, 816] & - \\
					Slow cooling period (s)		& [251, 4800] & [222, 3515] & - & - & [581, 2600] & [591, 2314] & - & - & [888, 1560] & [900, 1385] & - & - \\
					\bottomrule
			\end{tabular}}
		\end{threeparttable}
	\end{table*}
 
    The battery capacity loss, battery SoC, and battery temperature results under NYCC, SC03, and US06 are presented in Fig. \ref{fig13_comparison}, as marked with different colors. The proposed rule-based cooling strategy consists of three stages: fast cooling, slow cooling, and temperature maintenance, for each driving cycle. As shown in Fig. \ref{fig13_comparison} (a) and (b), the performance of the rule-based strategy is similar to that of DP, as demonstrated by the similarity in battery capacity loss and SoC values. In contrast, MPC has the highest battery capacity loss and lowest battery SoC under different driving cycles, despite consuming more cooling energy. As indicated in Fig. \ref{fig13_comparison} (c), the final battery temperatures can be well maintained around 25$^{\circ}$C by all three cooling methods. However, the enlarged figure shows that the cooling process of the proposed rule-based strategy is also similar to that of DP (almost overlap), while MPC cools the batter very fast ($P_{comp}$ is around 3000W and $T_{bat}$ reaches 25$^{\circ}$C before 1000s for all three driving cycles) and then stops cooling until the battery temperature rises, causing a jagged temperature curve of MPC in the temperature-maintaining stage.
 
	A numerical comparison between the proposed rule-based strategy, DP method, MPC, and no-cooling (as a benchmark) cases is provided, as summarized in Table \ref{tab6}.  For battery capacity loss, the proposed rule-based strategy achieves near-optimal performance with only a 2.46\% higher battery capacity loss compared to DP. Note that the rule-based strategy has the same final battery capacity loss as DP under US06, i.e., 0.0380, while it uses more battery energy. When compared with MPC, the proposed strategy reduces battery degradation by 1.8-3.8\% and saves 0.55-6.34\% consumed SoC. When compared with the no-cooling case, the proposed strategy reduces battery degradation by 18.83\%, 18.75\%, and 19.83\% under NYCC, SC03, and US06, respectively, while the battery SoC only reduces by 1.98, 1.44, and 2.55, respectively (the consumed SoC is 2.35\%/2.14\%/2.84\% more than the no-cooling case under NYCC/SC03/US06). Since the increased energy consumption is minor, the driving range of EVs is barely reduced.
	
	\subsection{Comparative results of driving economy and battery life\label{economy}}

        \begin{figure*}[!t]
		\centering
		\includegraphics[width=16cm]{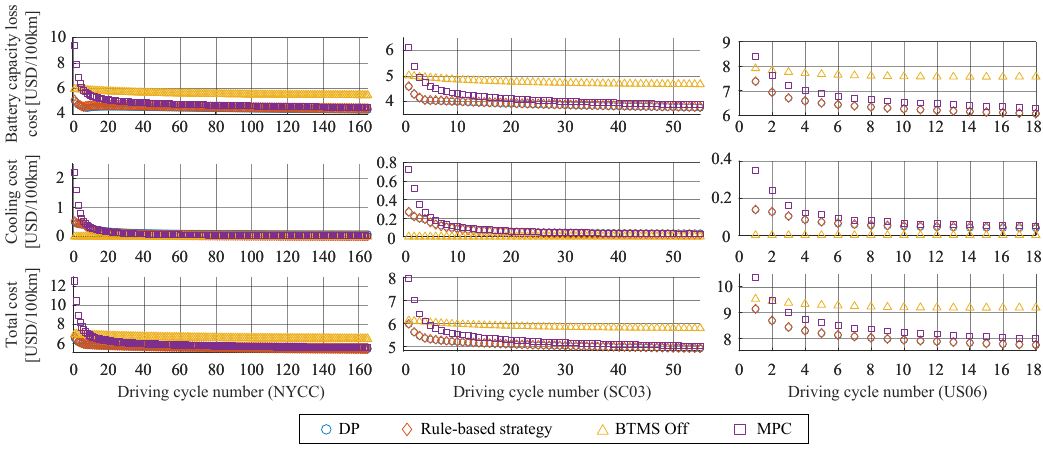}
		\caption{Economy comparison of DP, rule-based strategy, MPC, and BTMS Off cases under different driving conditions and different driving cycle numbers (NYCC, SC03, and US06).}
		\label{fig14_economy}
	\end{figure*}

        \begin{figure*}[!h]
		\centering
		\includegraphics[width=16cm]{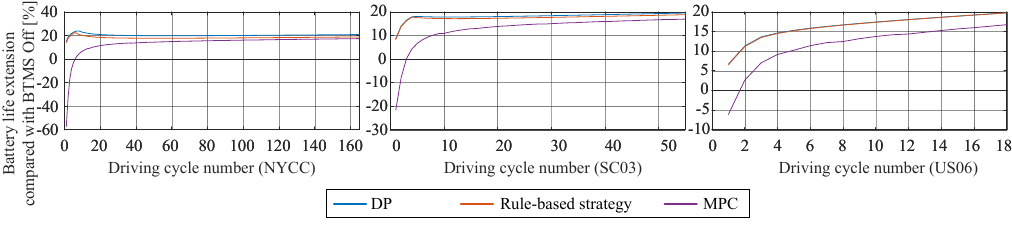}
		\caption{Battery life extension (compared with BTMS off) of DP, rule-based strategy, and MPC cases with different driving cycle numbers under different driving conditions (NYCC, SC03, and US06).}
		\label{fig15_life_extension}
	\end{figure*}
	
	The impact of BTMS on the driving economy under different driving distances is further quantified in this section. Fig. \ref{fig14_economy} presents the detailed driving economy results of DP, rule-based strategy, MPC, and the no-cooling case with different driving cycle numbers. Although the reductions of total cost and battery degradation cost are not very significant when the driving cycle number is 1 (i.e., short trips), DP and rule-based strategy still provide a lower total cost because of the reduced battery degradation, meaning that it is worth cooling the battery even for short trips. Since the battery cooling cost reduces over time, battery cooling becomes increasingly necessary over a longer trip, when compared to the no-cooling strategy, as the accumulated reduction of battery degradation is significant. In addition, as shown in Fig. \ref{fig14_economy}, the proposed rule-based strategy achieves similar results when compared to DP regarding the battery capacity loss cost, battery cooling cost, and total cost. The results are almost overlapped, indicating that the rule-based strategy can mimic DP very well under different types of driving cycles. It can be found that regardless of the adopted cooling strategy, cooling down the battery can reduce both the battery capacity loss cost and the total cost when compared to the no-cooling case. However, the MPC suffers from significantly higher driving costs at the beginning (i.e., the first 6/3/1 driving cycles under NYCC/SC03/US06), which means the MPC is not cost-effective for short-term driving. Note that in real-world applications, the prediction uncertainties can further deteriorate the MPC performance.

        The battery life extension results under different cooling strategies, driving conditions, and distances are presented in Fig. \ref{fig15_life_extension}. The proposed rule-based strategy is quite close to DP since both strategies can extend the battery life from the very beginning. As the battery temperature stabilizes, an approximately 20\% battery life extension can be achieved. In contrast, due to the lack of long-term prediction information and the limitation of local optimization, the MPC cooling method will shorten the battery life for short-term driving. This highlights the superiority of the proposed method, which leverages the regenerative power to cool the battery, thereby prolonging battery life even in the battery cooling process.
        
	An interesting phenomenon in Fig. \ref{fig15_life_extension} is that for DP and rule-based strategies, as the driving cycle number increases, the battery life extension increases first and then decreases under NYCC (i.e., urban driving condition), while for SC03 (i.e., suburban driving condition), such a phenomenon is much less obvious, and the battery life extension result is monotonic for US06 (i.e., highway driving condition). Specifically, at the beginning of urban driving conditions, the battery life extension mainly comes from the reduction of accumulated charging current, rather than the temperature reduction, as most regenerative braking energy is utilized to cool the battery. When the BTMS enters the temperature-maintaining stage, the charging process is more frequent during regenerative braking since the compressor power is around 0. As a result, although the battery temperature is reduced to 25$^{\circ}$C, the battery degradation cost starts to increase. In contrast, under highway conditions, the battery life extension mainly comes from the temperature reduction, not the reduced battery charging current, as there are no aggressive and frequent braking actions. Thus, after the BTMS enters the temperature-maintaining stage (driving cycle number of 3 for DP and rule-based strategy, see the 1560s and 1385s in Table \ref{tab6}), the battery capacity loss cost can still decrease. The suburban driving condition stays somewhere between the above two driving conditions; therefore, the non-monotonic phenomena in the battery life extension still exist but is less pronounced.
	
	\subsection{Parameter sensitivity analysis}

        \begin{figure*}[!b]
		\centering
		\includegraphics[width=16cm]{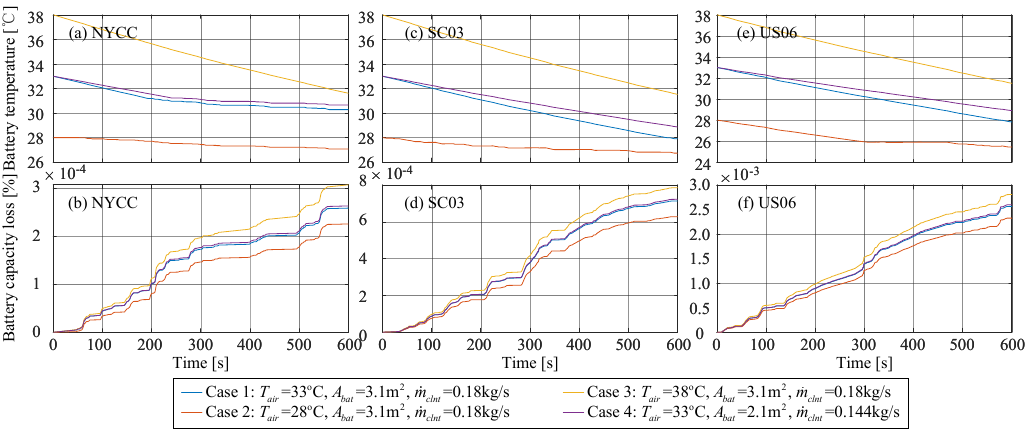}
		\caption{Cooling process of one driving cycle under different $T_{air}$, $\dot{m}_{clnt}$, and $A_{bat}$ parameter combinations. (a, c, e) for battery temperature under NYCC, SC03, and US06. (b, d, f) for battery capacity loss under NYCC, SC03, and US06.}
		\label{fig16_comparison_600s}
	\end{figure*}
	
	In addition to the tested parameters of the battery cooling (i.e., $T_{air}=33^{\circ}$C, $\dot{m}_{clnt}$=0.18kg/s, and $A_{bat}$=3.1m$^2$), it is necessary to study other parameter values to assess the robustness of the proposed cooling strategy, as various cooling systems can be adopted in EVs. Given that the control-oriented model of the BTMS (Eq. (\ref{eq:Q_cool})) is valid within the $T_{air}$ range of 26 - 40$^{\circ}$C and the $\dot{m}_{clnt}$ range of 0.144 - 0.216kg/s, we compare the following 4 parameter combinations in DP optimization, as listed in Table \ref{tab7}. Compared with Case 1, Case 2 and 3 are conducted to investigate different temperature conditions, and Case 4 is set to examine the results of different BTMS designs, e.g., with a worse cooling capability.
	
	\begin{table}[!h] 
		\caption{Four parameter combinations of $T_{air}$, $A_{bat}$, and $\dot{m}_{clnt}$}
		\label{tab7}
		\centering
		\begin{threeparttable}
			\resizebox{8cm}{!}{	
				\begin{tabular}{lcccl}
					\toprule
					Case	& $T_{air}$ ($^{\circ}$C)  & $\dot{m}_{clnt}$ (kg/s)	& $A_{bat}$ (m$^2$)	&	Note	\\
					\midrule
					Case 1		& 33	& 0.180  &  3.1	& benchmark 	\\
					Case 2		& 28 	& 0.180  &  3.1	& lower environment temperature 	\\
					Case 3		& 38	& 0.180  &  3.1	& higher environment temperature 	\\
					Case 4		& 33	& 0.144  &  2.1	& worse BTMS cooling capability 	\\
					\bottomrule
			\end{tabular}}
		\end{threeparttable}
	\end{table}
	
	In Fig. \ref{fig16_comparison_600s}, the battery temperature variation and battery capacity loss results of one driving cycle (600s) under the four parameter combinations are presented. Taking the NYCC (Fig. \ref{fig16_comparison_600s} (a, b)) as an example, it can be found that the battery capacity loss is mainly determined by $T_{air}$ (average battery temperature), rather than $A_{bat}$ and $\dot{m}_{clnt}$. Compared with Case 1 and 4, Case 3 has higher final battery temperatures ($\sim$0.96-1.34$^{\circ}$C) and therefore higher final battery capacity losses ($\sim$16-19\%), which also verifies the necessity of battery cooling in hot weather. BTMS parameters ($A_{bat}$ and $\dot{m}_{clnt}$) have a more limited impact on battery cooling, e.g., the battery temperature and capacity loss results of Case 1 and 4 are close, even though the cooling process of Case 4 is slower.
	
	Results of SC03 and US06 are also similar: the battery capacity loss is mainly determined by $T_{air}$ (average battery temperature), rather than $A_{bat}$ and $\dot{m}_{clnt}$. However, the difference lies in the fact that for SC03 and US06, the final battery temperature difference between Cases 1 and 4 is larger than that of NYCC. This is because the fast cooling stage of SC03 and US06 is much longer than NYCC (see Table \ref{tab5}), revealing that the BTMS parameters have a greater impact on the fast cooling stage (more heat exchange).
    To further investigate the effect of cooling parameters ($T_{air}$/$\dot{m}_{clnt}$/$A_{bat}$) on driving economy and battery degradation, the total cost reduction and battery degradation reduction compared with the no-cooling case are listed in Table \ref{tab8}. For different driving conditions, the short-trip and long-trip values denote the results of one driving cycle (600s) and the maximum driving cycle (165/55/18 for NYCC/SC03/US06), respectively.

        \begin{table*}[!h] 
		\caption{Total cost reduction and battery degradation reduction compared with the no-cooling case under different parameter combinations ($T_{air}$/$\dot{m}_{clnt}$/$A_{bat}$) and different driving cycles (NYCC/SC03/US06).}
		\label{tab8}
		\centering
		\begin{threeparttable}
			\resizebox{12cm}{!}{	
				\begin{tabular}{cccccccc}
					\toprule
					\multirow{2}{*}{Index} & \multirow{2}{*}{Case} & \multicolumn{2}{c}{NYCC} & \multicolumn{2}{c}{SC03} & \multicolumn{2}{c}{US06}	\\
					& & short-trip & long-trip & short-trip & long-trip & short-trip & long-trip	\\
					\midrule
					& Case 1 & 0.41 & 1.06 & 0.17 & 0.88 & 0.39 & 1.46 \\
					Total cost reduction & Case 2 & 0.53 & 0.52 & 0.28 & 0.44 & 0.31 & 0.82 \\
					(USD/100km) & Case 3 & -0.06 & 1.65 & 0.27 & 1.37 & 0.54 & 2.15 \\
					& Case 4 & 0.30 & 1.06 & 0.11 & 0.88 & 0.30 & 1.45 \\
					\midrule
					& Case 1 & 15.17 & 20.86 & 8.61 & 19.58 & 6.70 & 19.79 \\
					Battery degradation & Case 2 & 17.12 & 12.17 & 10.08 & 11.14 & 6.02 & 12.44 \\
					reduction (\%) & Case 3 & 9.74 & 28.62 & 9.66 & 27.18 & 7.78 & 26.37 \\
					  & Case 4 & 13.68 & 20.94 & 7.40 & 19.61 & 5.50 & 19.72 \\
					\bottomrule		
			\end{tabular}}
		\end{threeparttable}
	\end{table*}

    The effect of $\dot{m}_{clnt}$ and $A_{bat}$ can be found in Cases 1 and 4 ($T_{air} = 33^{\circ}$C). For different driving cycles, even though the total cost reduction and battery degradation reduction of Case 4 is lower than Case 1 under short-trip, the results will gradually converge as driving distance increases (long-trip). This also indicates that $\dot{m}_{clnt}$ and $A_{bat}$ have limited impact on driving economy and battery degradation. Since for long-trip driving (enough time to cool $T_{bat}$ to 25$^{\circ}$C), $T_{bat}$ can be well maintained with different $\dot{m}_{clnt}$/$A_{bat}$ parameters.
    In contrast, the environment temperature $T_{air}$ has a more significant effect on the driving economy and battery degradation. By comparing Cases 1, 2, and 3, it can be found that higher temperatures lead to greater reductions in both total cost and battery degradation during long trips. However, for short trips in urban areas (600s, $\sim$2km), battery cooling can increase the total cost in Case 3, even though the battery degradation is still reduced. This is because (1) the fast cooling process will take a longer time since the environmental temperature is high, and thus the cooling cost is higher than the reduced battery degradation cost, and (2) the cooling cost in short-term urban driving conditions is higher than that in suburban and highway conditions due to larger proportion of fast cooling power to total load power and the lower BTMS efficiency (see Eq. (\ref{eq:m_air})). Note that when the driving cycle number is 2 (NYCC, 1200s, $\sim$4km), there is only a limited cost reduction of 0.67USD/100km for Case 3. Due to this very limited cost reduction under short-trip urban driving, battery cooling is unnecessary. However, for long trips in hot weather, such as those experienced by Taxis \cite{niu2022optimal}, battery cooling is essential.
	
    From the analysis above, it can be found that there is no significant difference in the driving economy and battery degradation under different cooling systems. Recommendations regarding battery cooling are then given as follows: (1) The driving economy is mainly influenced by the driving distance and environment temperature, and battery cooling is more necessary under long-trip driving and high environment temperature; (2) Battery cooling will provide more total cost reduction under higher temperature environment; and (3) Battery cooling is essential for long-trip driving regardless of the driving conditions.
	
	\section{Conclusion \label{conclusion}}
	
	This paper investigates the battery cooling optimization of passenger EVs in hot weather. An electrical-thermal-aging model for commercial $\rm LiFePO_4$ batteries is adopted, and a control-oriented dynamic model of an onboard BTMS is established to characterize the relationship between the compressor power and the battery cooling rate under various operation conditions. Then the DP algorithm is introduced to simultaneously minimize the battery degradation cost and the battery cooling cost. By analyzing the DP results, an online near-optimal battery cooling strategy is proposed, which contains a fast cooling stage, a slow cooling stage, and a temperature-maintaining stage. Simulation results show that the rule-based strategy can achieve near-optimal performance when compared to DP, e.g., the battery degradation difference between the two strategies is less than 3\%.
	
	Detailed comparison with existing MPC and no-cooling cases, parameter sensitivity analysis, and driving/cooling economy analysis are conducted under urban, suburban, and highway driving conditions. Results show that (1) The proposed online cooling strategy can significantly reduce battery degradation without obviously reducing driving range; (2) Different BTMS parameters can influence the cooling speed but will not have a significant impact on driving economy; (3) The proposed BTMS has a better performance when the environment temperature increases and the trip becomes longer. The key finding of this study is that using as much regenerative energy as possible to cool the battery pack can dramatically improve the driving economy and reduce battery degradation. 
    
	\section*{Acknowledgment}
	
	This work was supported in part by the National Science Foundation of China (Grant No. 52272339) and Hunan ProvincialNatural Science Foundation (Grant No. 2021JJ30828), China Railway Signal and Communication Corporation Limited Research Project (Grant No. 2300-K1200035), and China Shenhua Heavy Freight Train Group Operation Control System Technology Project. The first author is supported by China Scholarship Council.
	
	Special thanks to the support by Magna PT Powertrain (Shanghai) Co., Ltd., for providing KULI technical service. Special thanks to Seho Park from the Pusan National University, South Korea for providing help with BTMS modeling, and Jiahao Huang from Huawei Technologies Co., Ltd., China for providing assistance with software.
	
	\section*{Appendix: BTMS orthogonal experiment and model parameters \label{Appendix}}
	
	To develop a BTMS model from the orthogonal experiment, this paper selects an onboard air-conditioning system officially supplied by Magna, and this model is built into KULI software which is well validated by experiments \cite{li2021innovative, qi2007analysis}. In this model, R-134a is adopted as the refrigerant for the air-conditioning loop, and 50\% ethylene glycol-water mixture is selected as the coolant for cooling the battery pack. According to existing studies \cite{wang2022cooling, wang2020eco}, the inlet temperature of the condenser, the air mass flow rate of the condenser, the outlet temperature of the battery coolant, the mass flow rate of coolant, and the rotation speed of the compressor are chosen as the main parameters to design the orthogonal experiment. To simplify the complex model, the following assumptions are adopted.

        \setcounter{figure}{0}
	\renewcommand{\thefigure}{\Alph{figure}}
	\begin{figure}[!h]
		\centering
		\includegraphics[width=8cm]{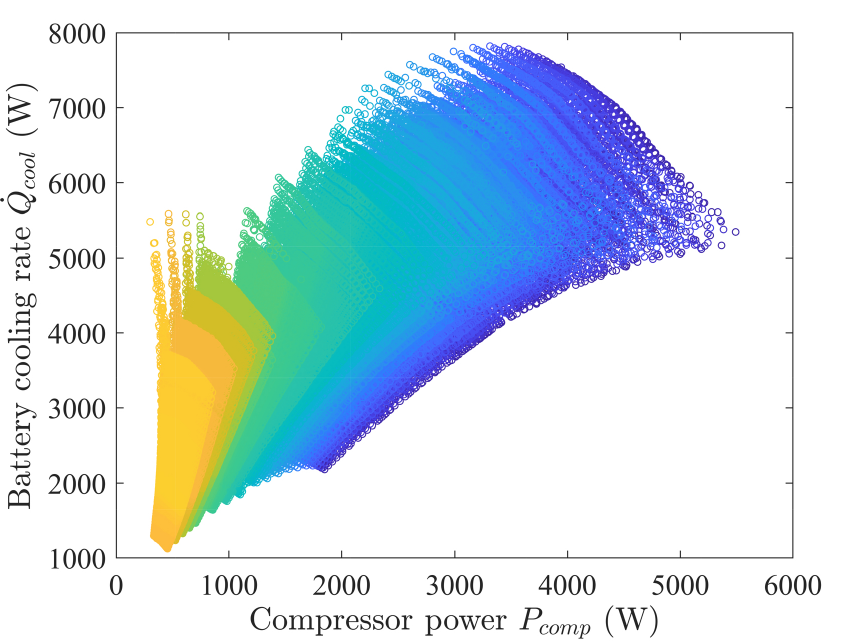}
		\caption{Simulation results: relationship between cooling rate $\dot{Q}_{cool}$ and compressor power $P_{comp}$.}
		\label{apdx_fig:Simdata}
	\end{figure}
    
	\begin{enumerate}
		\item The cooling fan for enhancing heat exchange is set to supply a constant air mass flow.
		\item All cooling capacity is used for the battery pack, i.e., the impact of the EV cabin is not taken into account.
	\end{enumerate}
	
	Therefore, experiment variables include the outlet temperature of coolant $T_{clnt,out}$, volumetric flow rate of coolant $\dot{q}_{clnt}$, ambient air temperature $T_{air}$, vehicle speed $v$ (affects inlet air mass flow rate of condenser, the total air mass flow $\dot{m}_{air}$ consists of air supplied by the EV velocity and fan), and electrical power of the compressor $P_{comp}$. The target variable is set as battery cooling rate $\dot{Q}_{cool}$. See Table \ref{tab_range}. 
	
	The relationship between the cooling rate and compressor power in the experiment is shown in Fig. \ref{apdx_fig:Simdata}. Note that it is almost impossible to fit a high-precision and accurate correlation formula with all data directly. Therefore, we divide data into regions and then fit them sequentially, where $T_{air}$ and $\dot{q}_{clnt}$ are selected as division basis. The BTMS model (Eq. (\ref{eq:Q_cool})) comprehensively considered compressor electrical power, ambient temperature, the outlet temperature of the coolant, air mass flow rate, and mass flow rate of the coolant. Detailed data of the six coefficients, i.e., Fig. \ref{fig4_lamda}, is attached.
	
	\bibliographystyle{elsarticle-num}
	\bibliography{Reference}
	
\end{document}